\documentclass[aps,pra,twocolumn,showpacs,superscriptaddress]{revtex4-2}
\usepackage[english]{babel} %%% 'french', 'german', 'spanish', 'danish', etc.
\usepackage{amssymb}
\usepackage{amsmath}
\usepackage{txfonts}
\usepackage{mathdots}
\usepackage[classicReIm]{kpfonts}
\usepackage[dvips]{graphicx}
\usepackage{epsfig}
\usepackage{graphicx}
\usepackage{array}
\usepackage{amssymb}
\usepackage{amsfonts}
\usepackage{amsmath}
\usepackage{mathrsfs}
\usepackage{color}
\usepackage{booktabs}
\usepackage{threeparttable}
\usepackage{multirow}
\usepackage{subfigure}
\usepackage{epsfig}
\usepackage{threeparttable}
\usepackage{chngpage}
\usepackage{float}
\usepackage{color}
\usepackage{bm}
\usepackage{graphicx}
\usepackage[toc]{appendix}

\begin{document}

\title{Strong and long-range radiative interaction between resonant transitions}

\author{Lei Ying} \email{leiying@zju.edu.cn}
\affiliation{Interdisciplinary Center for Quantum Information, State Key Laboratory of Modern Optical Instrumentation, and Zhejiang Province Key Laboratory of Quantum Technology and Device, Department of Physics, Zhejiang University, Hangzhou 310027, China}
\affiliation{Department of Electrical and Computer Engineering, University of Wisconsin, Madison, Wisconsin 53706, USA}

\author{Michael S. Mattei}
\affiliation{Department of Chemistry, University of Wisconsin, Madison, Wisconsin 53706, USA}

\author{Boyuan Liu}
\affiliation{Department of Electrical and Computer Engineering, University of Wisconsin, Madison, Wisconsin 53706, USA}

\author{Shi-Yao Zhu}
\affiliation{Interdisciplinary Center for Quantum Information, State Key Laboratory of Modern Optical Instrumentation, and Zhejiang Province Key Laboratory of Quantum Technology and Device, Department of Physics, Zhejiang University, Hangzhou 310027, China}

\author{Randall H. Goldsmith}
\affiliation{Department of Chemistry, University of Wisconsin, Madison, Wisconsin 53706, USA}

\author{Zongfu Yu} \email{zyu54@wisc.edu}
\affiliation{Department of Electrical and Computer Engineering, University of Wisconsin, Madison, Wisconsin 53706, USA}

\date{\today}

\begin{abstract}
Enhancing the radiative interaction between quantum electronic transitions is of general interest. There are two important properties of radiative interaction: the range and the strength. There has been a trade-off between the range and the strength observed in the literature. Such apparent trade-off arises from the dispersion relation of photonic environments. A general recipe is developed to overcome such trade-off and to simultaneously enhance the range and the strength of radiative interactions.
\end{abstract}

\maketitle

\section{Introduction}
The radiative interaction between electronic quantum transitions changes both the energy and lifetime of transitions. It leads to phenomena such as F$\ddot{\mathrm{o}}$rster energy transfer \cite{Cl:1995, GF:2017}, super-radiance \cite{SS:2009, SBF:2017,SMGF:2021}, and collective Lamb shift \cite{MSS:2014}. Because radiative coupling is short range, these phenomena can only be observed when the distance between transitions is not much longer than the wavelength of the light that mediates the interaction. Researchers have been interested in enhancing the strength and extending the range of radiative interactions, with applications in energy sciences and quantum technologies. While lots of progress has been made in the past decades, we observed that there has a been a trade-off in many methods studied in the literature \cite{FA:2013, MLE:2017, LE:2018, GSC:2015, SHK:2019, STU:2012, LR:2017, LLH:2007, Hu:2007, YH:2009, MS:2013, HMC:2013, VGG:2014, BMA:2016, CJ:2017, YZM:2019}. The trade-off occurs between the strength and the range: interactions with a long range are often weak in strength while strong interactions are very short range.

We first discuss how we define and quantify the \emph{range} $\ell$ and \emph{strength} $\Gamma_m$ of radiative interaction between quantum transitions. The radiative interaction includes two parts: $\Gamma(\mathbf{R})$ via propagating modes and $\Delta(\mathbf{R})$ via evanescent modes. Here we focus on the propagating modes because they are generally responsible for long-range interaction. We consider two quantum transitions in vacuum. Figure 1(b) plots the real photon part of radiative coupling coefficient as a function of the distance between two quantum transitions in vacuum (see Appendix~\ref{app:photonics} for detailed derivation on the radiative interaction), which describes the rate of energy exchange. The \emph{strength} is defined by the maximum value of the coupling, which is commonly realized when the distance is close to zero. The \emph{range} $\ell$ is defined by the distance at which the coupling decays to its half maximum value. For example, the interaction strength and range in vacuum are the spontaneous emission rate $\Gamma_0$ and $\ell=\ell_0\approx 0.6 \lambda_0$, respectively.

\begin{figure*}[htbp]
\centering\includegraphics[width=5in]{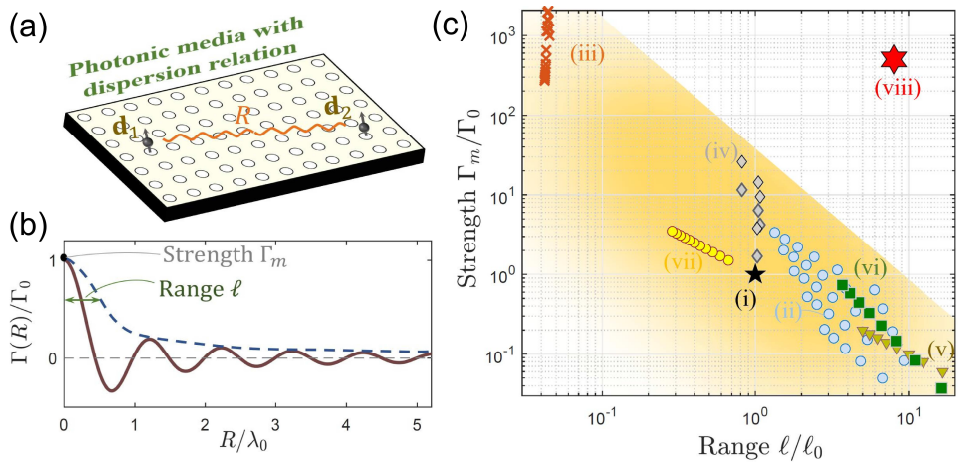}
\caption{(a) Schematic of two quantum transitions with separation distance $R$ embedded in arbitrary two- and three-dimensional photonic media with an explicit dispersion relation. (b) The interaction between two quantum transitions in vacuum. The strength defines the maximum interaction while the range defines the distance when it decays to half of its maximum. (c) Many existing works exhibit a trade-off between the interaction strength and range. The interaction has been studied in (i) Free space \cite{Le:1970}, (ii) Photonic crystal bandedge  \cite{JW:1991, NPR:2018}, (iii) Hyperbolic material \cite{CJ:2017}, (iv) Plasmonic plate \cite{ZLL:2011}, (v) index-near-zero material \cite{MLE:2017}, (vi) Weyl photonic crystals/lattices \cite{YZM:2019, GGJ:2020}, (vii) uniform medium , and (viii) this work, respectively.}
\label{fig:tradeoff}
\end{figure*}

Radiative interaction via real photon exchange is mediated through optical modes in the continuum. A general strategy to enhance radiative interaction strength and range is to place transitions in engineered photonic environments, where the optical modes can be drastically different from those in vacuum. This strategy was used in many interesting works that enhance the range \cite{FA:2013, MLE:2017, LE:2018, GSC:2015, SHK:2019, SBF:2017, LR:2017} or the strength \cite{BMA:2016, CJ:2017,HCLK:1998,GHYH:2014}. For example, zero-index materials can extend the range of coupling \cite{FA:2013, MLE:2017, LE:2018, GSC:2015, SHK:2019} to $F\ell_0$, where the enhancement factor $F$ can be a few hundreds. However, the strength $\Gamma$ decrease by a factor of $F^2$. The greater the range is, the weaker the strength becomes. Another interesting example exploits local plasmonic resonance \cite{ZLL:2011,GMMM:2011,ZYWP:2012}. The strength can be enhanced. However, the range becomes shorter \cite{ZLL:2011} due to the fast decay of plasmonic modes. There are other interesting works that use near fields \cite{AL:2002,Hu:2007, DHHG:2015,GHCC:2015, NPR:2018} or far fields around the Weyl point \cite{YZM:2019} in photonic crystals. When extending the cases to three-dimensional space, they all exhibit similar trade-off between range and strength. The only exception is one-dimensional space, such as in a waveguide or along the one particular direction of photonic structures where the interaction range can be extended without compromising the strength \cite{ZB:2013,LR:2017,SBF:2017,SMGF:2021}.

Figure 1(c) surveys recent works in which a trade-off between the interaction range and the strength can be observed. Although the specific mechanism for each case could vary, a common physics emerges in many of these cases. By identifying what contributes to the apparent trade-off, we show how to simultaneously enhance the strength and the range. The red hexagon in Fig. 1(c) for example has a $500$- and $8$-time enhancement of the strength and the range, respectively. Such enhancement could be very useful for quantum systems that use radiative interaction as the main mechanism for energy or information transduction.

\section{Long-range interaction}
We choose Dirac point because its isosurface can change rapidly while maintaining a relatively constant group velocity, which helps to isolate the effect of the contribution from isosurface.
The Hamiltonian near a Dirac point is given by $H_\mathrm{D}=v(k_x \sigma_x+k_y \sigma_y)$, where $\sigma_{x(y)}$ is the Pauli matrices and $v$ is the isotropic group velocity. This Hamiltonian results in a linear dispersion relationship as $\omega=v|\mathbf{k}-\mathbf{k}_\mathrm{c}|$, where $\mathbf{k}_\mathrm{c}$ is the location of the Dirac point in momentum space.

We study the interaction between two identical transitions placed inside the photonic crystal with the Dirac dispersion~\cite{PM:1991, SBB:2007} as shown in Fig. 1(a). We show three cases with their transition frequencies labeled in Fig. 2(c). From the point iii to point i in Fig. 2(c), the frequency gradually approaches the Dirac point. The radiative interaction as a function of the distance is shown in Fig. 2(d). The calculation is available in Appendix~\ref{app:num}. When the transition frequency is far away from the Dirac point, i.e. frequency-iii in Fig. 2(c), and the blue curve in Fig. 2(d), the strength and the range are not significantly different from those in vacuum. As we move the transition frequency close to the Dirac point, we see enhanced range at frequency-ii and frequency-i. In the case very close to the Dirac point (green curve in Fig. 2(d), the interaction decays slowly as the distance between the transitions increases. Conversely, the strength is weakened as the transition frequency approaches the Dirac point. Note that the green curve (point i) is displayed with a $9\times$ multiplication factor in Fig. 2(d). We summarize the strength and range of three cases in Fig. 2(b), and the trade-off is clearly observed: while the strength increases, the range decreases.

The trade-off observed here is quite general. We now discuss its origin. We consider a general continuum with a dispersion relationship $\omega=\omega(\mathbf{k})$, where $\mathbf{k}$ is the wavevector. We focus on optical modes around the transition frequency $\omega_0$ because they are the most important for mediating the radiative interaction. In the momentum space, these modes form an isosurface defined by $\mathbf{k}=\mathbf{k}(\omega_0)$. For example, the isosurface in vacuum is a sphere with a radius of $k=\omega_0/c$. In engineered photonic environments, the shape and size of isosurface can be quite different. We recently showed that the isosurface plays a significant role in radiative interactions \cite{YZM:2019, ZYL:2017}.

\begin{figure*}[htbp]
\centering\includegraphics[width=4.8in]{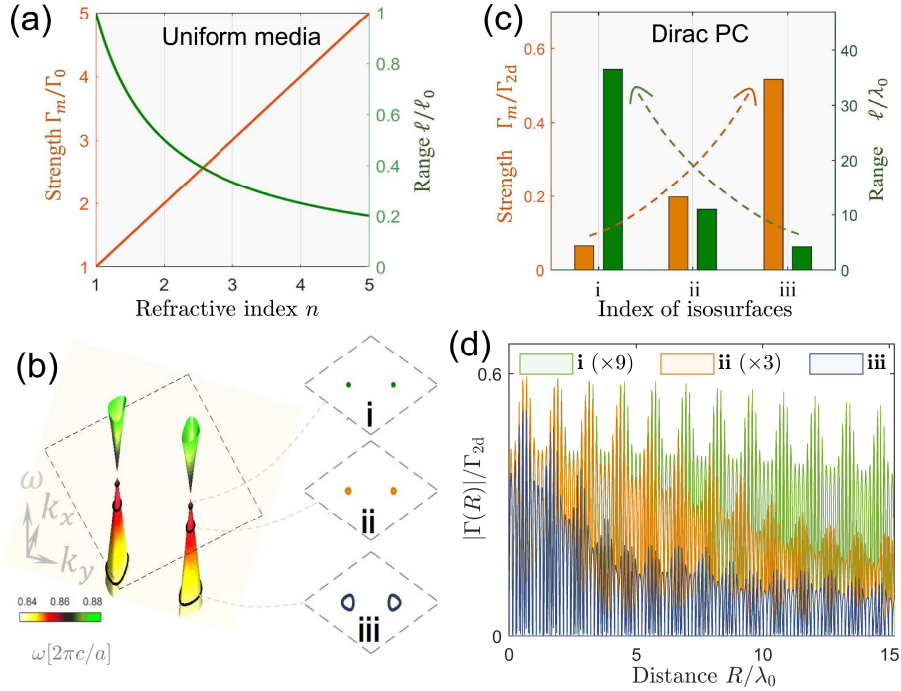}
\caption{(a) The tradeoff relationship between the interaction strength and the interaction range with the variation of the refractive index $n$ in the uniform media.
(b) Left: The dispersion relationship of the Dirac system. Right: Isosurfaces in momentum space with frequencies $\omega=0.86,0.855$, and $0.84\times 2\pi c/a$, respectively. Here, $a$ is the lattice constant of the triangular slab and $c$ is the speed of light. The Dirac-like dispersion relation is realized by a triangular photonic slab with the ratio between the radius of the hole and inter-hole distance $r/b=0.26$ and the refractive index is $2.83$.
(c) Comparison of the interaction strengths and the interaction ranges with isosurfaces i, ii, iii in (b). $\Gamma_\mathrm{2d}$ is the decay rate in two-dimensional space.
(d) The interaction via propagating modes as a function of inter-distance $R$ between quantum resonant transitions with isosurfaces i, ii, iii in (b). The dipole orientations are fixed at $[1,0]$.}
\label{fig:dirac}
\end{figure*}

\begin{figure*}[htbp] 
\centering\includegraphics[width=7in]{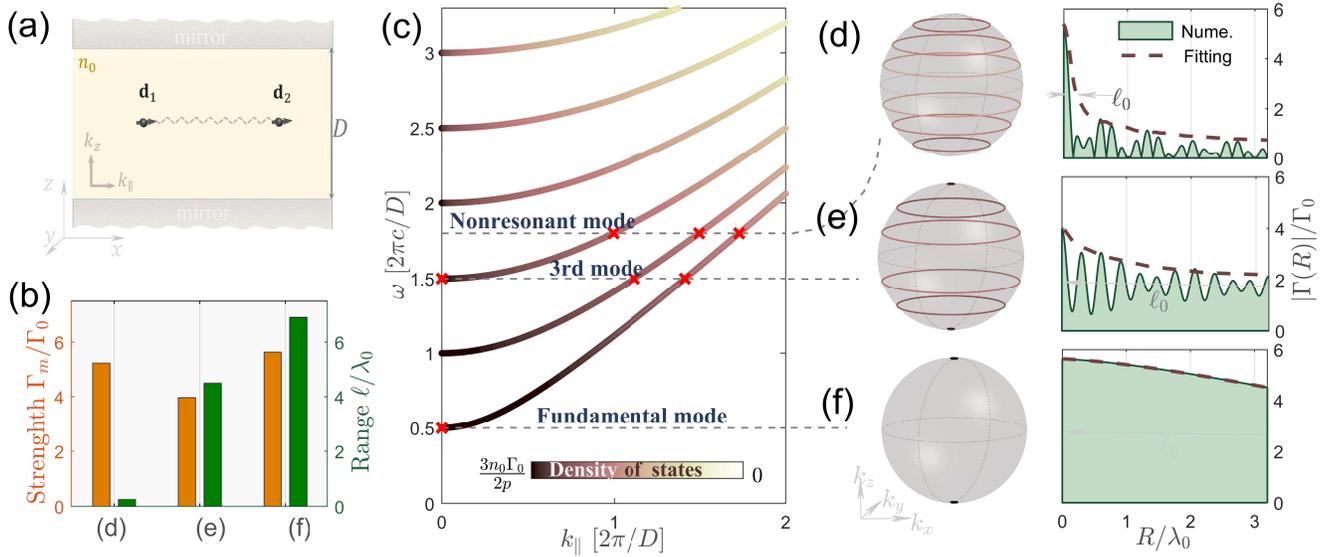}
\caption{(a) Schematic of a Fabry-P\'{e}rot cavity. Two quantum transitions are located in the high-index layer. (b) Interaction range and strength between two quantum transitions as a function of different cavity modes. Here we assume that the mirrors have a reflectance of $0.9998$. (c) The dispersion relation of an ideal Fabry-P\'{e}rot cavity. The color denotes the values of density of states for each mode. The reduced polarization factor is $p=d_1d_2\omega/16\pi^2 \varepsilon_0$.  (d-f) Left column: iso-frequency contours at three different frequencies. Right column: radiative interaction as a function of separation distance between resonant dipolar transitions at three different frequencies. The decay range $\ell$ is defined by the distance where the envelop of the interaction decays to half of its maximum. } 
\label{fig:cavity}
\end{figure*}

The radiative interaction can be written in the following form \cite{Le:1970,YZM:2019}
\begin{equation}\label{eq:eq1}
\Gamma(\omega,\mathbf{R})=\oiint_{\Sigma(\omega)}\frac{p_\mathbf{k}}{v_\mathbf{k}} e^{i\mathbf{k}\cdot\mathbf{R}} d\sigma   ,
\end{equation}
where the polarization factor $p_\mathbf{k}=\omega[\mathbf{d}_1 \cdot \mathbf{E}_\mathbf{k} (\mathbf{r}_1 )]^* [\mathbf{d}_2 \cdot \mathbf{E}_\mathbf{k}(\mathbf{r}_2 )]/16\pi ^2 \epsilon$ and $v_k=\sqrt{v_x^2+v_y^2+v_z^2}$  is the group velocity. The integration is performed over the isosurface $\Sigma(\omega)$. $\mathbf{E}_\mathbf{k}$ is the electric field of mode, $\mathbf{k}$. $\mathbf{d}_{1,2}$  are the dipole moments of the two transitions, $\mathbf{R}=\mathbf{r}_2-\mathbf{r}_1$ is the distance between the two transitions, and $\epsilon=n^2\epsilon_0$ is the dielectric constant and $n$ is the refractive index.

Now we are ready to explain the range of interaction. The term $e^{i\mathbf{k}\cdot \mathbf{R}}$ is a fast-oscillating term. The integration of this fast-oscillating term $e^{i\mathbf{k}\cdot \mathbf{R}}$ cancels when $\mathbf{R}$, the distance between the two transitions, is large. This is why the interaction vanishes for large $R$. In addition to $R$, the photonic environment also plays an important role through the shape of the isosurface $\Sigma(\omega)$. To see this impact, we fix a distance $\mathbf{R}$ between two transitions and consider photonic environments with different $\Sigma(\omega)$. If we have a large isosurface, the integration of $e^{i\mathbf{k}\cdot \mathbf{R}}$ cancels more compared to a small isosurface. As a result, a large isosurface leads to a shorter-range interaction than a small isosurface. This is what we observed in Fig. 2(b) and 2(c). The isosurface, which is an iso-contour in this two-dimensional space, is smaller around Dirac point, leading to a long interaction range. Away from the Dirac point, the isosurface is larger, resulting in a short interaction range.

Now we understand one aspect of the interaction-the range. We can examine the other side of the trade-off. The size of isosurface has a quite different impact on interaction strength. Larger isosurface generally leads to a greater interaction strength. The maximum interaction strength is obtained at $R = 0$  when $e^{i\mathbf{k}\cdot \mathbf{R}}$ becomes unity. A large isosurface has a large integration area, and thus a greater integration value. This relationship can also be understood using the optical density of states (DOS), which is proportional to the size of isosurface. The larger the DOS, the greater the spontaneous emission rate, which is directly related to the interaction strength.

Now the trade-off in Fig. 2(d) can be easily understood. Close to the Dirac point, the isosurfaces become smaller and smaller (Fig. 2(b)). Correspondingly, the range of interaction increases. However, smaller isosurfaces reduce the density of states, which decreases the spontaneous emission rate and the interaction strength.

{The understanding of interaction above is general for a photonic environment with an explicit dispersion relationship.} The trade-off caused by isosurfaces is built into the mathematic integration Eq.~(\ref{eq:eq1}) that determines the interaction. To mitigate this trade-off, we have to look into other physical mechanisms that contribute to the interaction: the first is the local field effect in the term $p_\mathbf{k}$, and the other is the group velocity in the term $v_\mathbf{k}$. By carefully balancing the isosurface, local field, and group velocity in judiciously designed photonic environments, we can overcome the apparent trade-off to realize simultaneous enhancement of interaction strength and range.

Our strategy starts by considering a photonic environment with a small isosurface because this is necessary to obtain a long range. However, it comes with a weak interaction strength. We can then use strong local fields and small group velocity to offset the negative impact from the small isosurface. The result is simultaneous enhancement of strength and range by many orders of magnitude. This strategy can be applied to many different types of engineered photonic media. Here we choose one of the simplest implementations in a Fabry-P\'{e}rot cavity~\cite{KZS:1995,VGKB:2018} to illustrate this method.

The cavity structure consisting two mirrors is shown in Fig. 3(a). The mirrors are extremely large and restrict the available modes and reduces the spherical isosurface in vacuum to iso-contours such as those rings and points shown in Fig. 3(c)-(f). Here we are particularly interested in the frequency region around the fundamental mode $\omega=\pi c/n_0 D$, where $D$ is the cavity length and $n_0$ is the refractive index of material inside the cavity. The isosurface of the fundamental mode reduces to two points (Fig. 3(f)). In comparison, high-order resonant modes consist of ring-shape isosurfaces (Fig. 3(d,e)). The diminishing isosurface at the fundamental mode is ideal for extending the interaction range. Unlike Dirac points and Weyl points, here the group velocity is zero in the $x$-$y$ plane, which helps to off-set the negative impact of small isosurface on interaction strength.

We perform direct calculation of the resonant dipole-dipole interaction inside a Fabry-P\'{e}rot cavity (see more details in Appendix~\ref{app:theory}). Fig. 3(b) shows the strength and range comparing the fundamental Fabry-P\'{e}rot modes to two other modes in the cavity. We can see the interaction range is longest for the fundamental mode because its smallest isosurface. Here we managed to maintain the strength of interaction because of the low group velocity.

\begin{figure}[htbp] 
\centering\includegraphics[width=\linewidth]{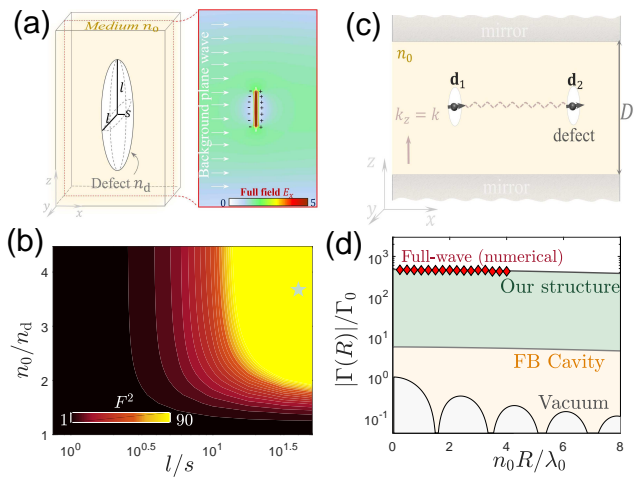}
\caption{(a) Left: The geometrical shape of low-index ellipsoidal defect with radiuses $l$ and $s$ along $y,z$ and $x$ axes, respectively. $n_\mathrm{d}$ and $n_0$ stand for the refractive indexes inside and outside the defect. Right: Electric field $E_x$ near and inside a low-index defect. ``+'' and ``-'' schematically represent the effective polarization. (b) The enhancement factor $F^2$ as a function of index ratio $n_0/n_\mathrm{d}$ and geometrical ratio $l/s$. (c) Schematic of dielectric structure for long-and-strong interaction. Two air defects (refractive index $n_\mathrm{d}$) are doped in a fundamental Fabry-P\'{e}rot cavity with high-index dielectric materials ($n_0$). (d) Absolution value or radiative interaction via propagating modes as a function of inter distance for cases of homogeneous dielectric media with refractive index $n_0$ (gray), fundamental-mode cavity (orange), and our proposed structure (green) in (c) based on theoretical estimation. Red diamonds denote the full-wave simulation results. The mirror reflectance of the cavity is $0.9985$ and the geometrical ratio is $l/s=40$. The enhancement factor of our proposal structure is marked as light green pentagram in (c). }
\label{fig:stronglong_structure}
\end{figure}

\section{Enhancement of interaction strength}
The fundamental mode in a Fabry-P\'{e}rot cavity provides a long interaction range while maintaining an interaction strength similar to that in vacuum. Next, we explore a second mechanism to greatly enhance the strength beyond that of vacuum using the local field effect \cite{bk:1960}. This effect occurs at the interface between two materials of different refractive indices when a small volume of the lower index material is embedded in the higher index material. Polarization charge accumulated at the interfaces creates enhanced local electric field \cite{bk:1960,ASL:2006,YRF:2010}, which can be used to enhance the field $\mathbf{E}_\mathbf{k} (\mathbf{r}_{1,2})$ in Eq. (\ref{eq:eq1}). To illustrate this effect, we consider an ellipsoidal defect of low index $n_\mathrm{d}$ in a material of high index $n_0$. The dimension of the ellipsoid is much smaller than wavelength (typically smaller than $\lambda_0/10$). Fig. 4(a) shows the field distribution under plane wave illumination with electric field polarized in $x$ direction. The intensity of the field inside is enhanced by $F^2=n_0^4  /[(1-r) n_0^2+r n_\mathrm{d}^2 ]^2$, where $r=(l^2 s/2) \int_0^\infty dq/[(q+l^2 )^2 (q+s^2 )^{1/2} ]$ is the geometrical factor that depends on the major radius $l$ and minor radius $s$ of the ellipsoid (see details in Appendix~\ref{app:enhancement}). Figure 4(b) shows the enhancement of the field intensity defects with different index contrast and geometrical ratio. With an index contrast around $3$, $90$ times enhancement can be realized for elongated defect shapes. Since it is small with a deep sub wavelength size, the defect does not significantly affect the modes in the continuum which are mostly determined by larger and wavelength-scale structures. The local field effect explores a completely different physics of enhancement than the group velocity and isosurface. It does not interfere with other enhancement mechanisms.

Next, we combine the effects of local field enhancement and the fundamental mode of a Fabry-P\'{e}rot cavity to demonstrate a case of long and strong interaction. The cavity is filled with a high-index material $n_0=3.67$. Inside the cavity, we have two small air holes as the low-index defects. We consider one resonant dipole inside each defect as shown in Fig. 4(c). The two dipoles are aligned to the short axis of the ellipsoidal defect. The radiative interaction between two quantum transitions is given by $\Gamma=(2\omega^2/\epsilon_0 c^2 ) \mathrm{Im}[\mathbf{d}_2\cdot \bar{\bar{\mathbf{G}}}(\mathbf{r}_2,\mathbf{r}_1;\omega)\cdot \mathbf{d}_1 ]$, where $\bar{\bar{\mathbf{G}}}(\mathbf{r}_2,\mathbf{r}_1;\omega)$ is the electric dyadic Green's function of the photonic environment. Then dyadic Green's function can be obtained numerically by a full-wave simulation shown in Appendix~\ref{app:num}.  The numerical results are marked as the red diamonds in Fig. 4 (d), which agree with our theoretical estimation in Appendix~\ref{app:enhancement}. For the cavity with a quality factor of $3.14\times10^4$, the same as the cavity in Fig. 3, the interaction strength can reach up to about $500\Gamma_0$. Meanwhile, the interaction range is $8$ times the range in vacuum. We also examine the results from non-Markovian calculation, which shows fairly consistent conclusions as the Markovian approximation used here. The details can be found in Appendix~\ref{app:dynamics}.
{The retardation effect is included in the full electrodynamic treatment~\cite{MK:1974}.}

\section{Discussion}

To conclude, we discuss the origin of the apparent trade-off between the interaction strength and interaction range. We analyze three main factors that contribute to radiative interaction: the isosurface of photonic environment, the local field, and the group velocity. The isosurface leads to the trade-off between the strength and the range of interaction. The local field and group velocity can be modified relatively independently. With this recognition, we developed a recipe for long and strong range interaction. We start with a small isosurface to guarantee a slow decay with a long range. Then we engineer the photonic environment to exploit the local field and the group velocity. It leads to long and strong interaction between two resonant transitions. It would be interesting to study a network of quantum transitions where they interact with each other strongly over extended distance, which could lead to quite different results from networks that only consider nearest neighbor interactions.

\section*{Acknowledgments}
This work was supported by the National Science Foundation (NSF) through the University of Wisconsin Materials Research Science and Engineering Center DMR-1720415. L. Y. was also supported by the National Key Research and Development Program of China 2019YFA0308100.

\appendix
\section{Theory of strong-and-long range interaction}\label{app:theory}

We consider the resonant dipole-dipole interaction between electric dipolar quantum transitions (QTs). The Hamiltonian of two identical QTs in an arbitrary photonic environment is given by
\begin{equation}\label{eq:hamiltonian_tot}
   \mathcal{H} = \mathcal{H}_{Q}+\mathcal{H}_{P}+\mathcal{H}_{I}.
\end{equation}
They are explicitly written as~\cite{bay1997atom} ($\hbar=1$)
\begin{equation}
\begin{split}
   \mathcal{H}_{Q} = & \omega_0 \left( \mathcal{S}_{1}^{\dag} \mathcal{S}_{1}  +  \mathcal{S}_{2}^{\dag} \mathcal{S}_{2}\right), \ \ \ \
   \mathcal{H}_{P} = \sum_{\alpha}\omega_\alpha \mathcal{A}_{\alpha}^{\dag} \mathcal{A}_{\alpha}, \\
   \mathcal{H}_{I} = & \sum_{\alpha}
   \Big[ g_{\alpha}(\mathbf{r}_1)
   \left(\mathcal{S}_{1}^{\dag} + \mathcal{S}_{1}\right) \mathcal{A}^\dagger_{\alpha}
+  g_{\alpha}(\mathbf{r}_2)
   \left(\mathcal{S}_{2}^{\dag} + \mathcal{S}_{2}\right) \mathcal{A}^\dagger_{\alpha} +\mathrm{h.c.} \Big],
\end{split}
\end{equation}
where $\omega_0$ and $\mathcal{S}_{1,2}^\dag (\mathcal{S}_{1,2})$ are the transition frequency and raising (lowering) operator of two-level QTs. $\omega_\alpha$ and $\mathcal{A}_{\alpha}^{\dag} (\mathcal{A}_{\alpha})$ are the frequency and creation (annihilation) operator of photon.
\begin{equation}
g_{\alpha}(\mathbf{r}_{1,2})=\sqrt{\frac{\omega_\alpha}{2 n_0^2 \epsilon_0 V_{P}} }     e^{-i \mathbf{k}\cdot \mathbf{r}_{1,2}}
\left[  \mathbf{d}_{1,2}\cdot \mathbf{E}_{\alpha}   \right]
\end{equation}
is the coupling between QTs and photonic mode $\alpha$, where the mode index $\alpha$ contains the information of photonic wavevector $\mathbf{k}$ and polarization $\mathbf{E}_\mathbf{k}$. $\mathbf{d}_{1,2}$ is the transition dipole moment of QTs. $V_P$ is the photon volume and $n_0$ is the refractive index of the photonic medium.

The transition probability from initial to final states is given by the Fermi's Golden rule
$ 2\pi / \hbar |\Sigma_{12} |^2
\delta\left( \mathcal{E}_\mathrm{F}-\mathcal{E}_\mathrm{I} \right)$,
where the transition matrix element $\mathcal{M}_\mathrm{F I}$ can describe the resonant dipole-dipole interaction between two QTs.
For the weak light-matter interaction, it can be written as the second-order form:
\begin{equation}
 \Sigma_{12}
= \langle \mathrm{F} | \mathcal{H}_{I} | \mathrm{I} \rangle
+ \sum_m \frac{ \langle \mathrm{F}| \mathcal{H}_{I} | m \rangle  \langle m | \mathcal{H}_{I} | \mathrm{I} \rangle   }{\mathcal{E}_\mathrm{I} -\mathcal{E}_m } + \cdots
\end{equation}
Here, $|\mathrm{I}\rangle = | e_1,g_2;0\rangle$ and $|\mathrm{F}\rangle = | g_1,e_2;0\rangle$ denote initial and final states, where`$e$' and `$g$' in the Dirac bracket notions represent excited and ground states, respectively, and the number `$0$' or `$1$' is the photon number in the photonic environment. The intermediate state $| m \rangle$ has two options: $|g_1,g_2;1_\alpha \rangle$ with energy $\mathcal{E}_{ m1}=\mathcal{E}_g^{(1)}+\mathcal{E}_g^{(2)}+\hbar \omega_\alpha$ and $| e_1,e_2;1_\alpha \rangle$ with energy $\mathcal{E}_{m2 }=\mathcal{E}_e^{(1)}+\mathcal{E}_e^{(2)}+\hbar \omega_\alpha$. The energy of the initial state is $\mathcal{E}_{\mathrm{I} }=\mathcal{E}_g^{(1)}+\mathcal{E}_g^{(2)}+\hbar \omega_\alpha$. Since two identical QTs are considered, we have $\mathcal{E}_e^{(1,2)}-\mathcal{E}_g^{(1,2)}=\hbar \omega_0$. Then, Eq. (3) can be written as
\begin{equation}
\Sigma_{12}=
\sum_{\alpha}
\Big[
 \frac{ g_{\alpha}^\ast\left(\mathbf{r}_1\right) g_{\alpha} \left(\mathbf{r}_2\right)   }{ \omega_\alpha -\omega_0 }
 +
\frac{ g_\alpha \left(\mathbf{r}_1\right) g_{\alpha}^\ast\left(\mathbf{r}_2\right) }{ \omega_\alpha + \omega_0 } \Big] .
\end{equation}

{\bf Analytical solutions of radiative interactions via propagating modes in a cavity ---}
We consider a Fabry$-$P\'{e}rot with two PEC plates. The volume is given by
\begin{equation}
V_P=L_x L_y D,
\end{equation}
where $D$ is the thickness of the cavity and $L_{x, y} \rightarrow\infty$. Then, the summation over mode index $\alpha$ can be written as an integral:
\begin{equation}
\sum_\alpha \rightarrow \sum_{k_z}  \frac{L_x L_y}{(2\pi)^2}   \iint dk_x dk_y.
\end{equation}
 Also, utilizing the relationship \begin{equation}
 \frac{\omega}{\omega-\omega_0}dx= \mathbb{P} \frac{\omega}{\omega-\omega_0} + i\pi\delta(\omega-\omega_0) ,
\end{equation}
the interaction is written as
\begin{equation}\label{eq:interaction_expression}
\begin{split}
\Sigma_{12}(\omega,\mathbf{R}) =& \Delta + i\Gamma  \\
= & \sum_{k_z} \iint_{l_{k_z}} dk_x dk_y
\Bigg[  i\pi\delta(\omega-\omega_0) p_k \\
& + \mathbb{P}
\left(
\frac{e^{i\mathbf{k}\cdot \mathbf{R}}}{\omega-\omega_0} p_\mathbf{k}
+ \frac{e^{-i\mathbf{k}\cdot \mathbf{R}}}{\omega+\omega_0} p_\mathbf{k}^\ast
\right)
\Bigg],
\end{split}
\end{equation}
where $\mathbb{P}$ is the Cauchy principal value and $\mathbf{R}=\mathbf{r}_2 - \mathbf{r}_1$.
The polarization factor is written as
\begin{equation}
p_\mathbf{k} =  \frac{\omega}{8\pi^2 n_0^2 \epsilon_0 D} \left[ \mathbf{d}_1 \cdot \mathbf{E}_\mathbf{k}(\mathbf{r}_1) \right]^\ast  \left[ \mathbf{d}_2 \cdot \mathbf{E}_\mathbf{k}(\mathbf{r}_2) \right],
\end{equation}
where the dipole moment and (degenerate) electric polarizations of mode $\mathbf{k}$ can written as
\begin{equation}
\begin{split}
\mathbf{d}_{i} = & [d_{i,x} , d_{i,y}, d_{i,z}]^T     \\
 = &
 d_{i} \left[ \cos\varphi_{i} \cos \theta_{i} , \cos\varphi_{i} \sin \theta_{i}, \sin \varphi_{i} \right]^T , \\
\mathbf{E}_\mathbf{k} = &
\left[ \sin \varphi_k \cos \theta, \cos \varphi_k \cos \theta_k, \sin \varphi_k \right]^T
 \\
+ &
 \left[ \sin \theta_k, \cos \theta_k, 0 \right]^T ,
\end{split}
\end{equation}
where $i=1,2$. The radiative interaction is given by
\begin{equation}
\Gamma(\omega_0,\mathbf{R}) =
\sum_{k_z}
e^{ik_z \Delta z }
\frac{\pi}{v_\parallel}
\oint_{ \mathcal{C}(\omega_0) } dl
p_{\mathbf{k}}e^{i\mathbf{k}_\parallel \cdot \mathbf{R}_\parallel},
\end{equation}
where $\mathbf{R}_{\parallel}=\left[ x_2-x_1, y_2-y_1\right]$, $\Delta z= z_2-z_1$ and $k_\parallel=\sqrt{k^2-k_z^2}$. $\mathcal{C(\omega)}$ is the circular isosurface with a same frequency $\omega$.
The group velocity in the $x-y$ plane is
\begin{equation}
v_\parallel= \partial\omega / \partial k_\parallel = c k_\parallel/n_0 k.
\end{equation}
Then, the radiative interaction is written as
\begin{equation}
\Gamma(\omega,\mathbf{R})
 =
\frac{d_1 d_2 \omega_0}{2\pi n_0^2 \epsilon_0 }
\sum_{k_z} \frac{ k_\parallel \cos{k_z \Delta z}}{v_\parallel D}
F\left( k_\parallel,R \right),
\end{equation}
where
\begin{equation}
 F\left( k_\parallel,R_\parallel \right) = 2\pi    \left[
 \frac{ J_1(k_\parallel R_\parallel)}{k_\parallel R_\parallel} \alpha_{12}
 - J_2(k_\parallel  R_\parallel)\beta_{12}
  \right],
\end{equation}
where
\begin{equation}
\begin{split}
\alpha_{12} = & \left[ 1+ \cos(\varphi_1 - \varphi_k) \cos(\varphi_2 - \varphi_k) \right] \cos(\theta_1 - \theta_2),  \\
\beta_{12} = & \cos \theta_1 \cos \theta_2 \cos(\varphi_1-\varphi_k) \cos(\varphi_2-\varphi_k)   \\
   & \ \ \ \ \ \ \ \ \ \ \ \ \ \ \ \ \ \ \ \ \ \ \ \ \ \ \ \ \ \ \ \ \ \  + \sin \theta_1 \sin \theta_2 .
\end{split}
\end{equation}
Then, we consider the fundamental-mode cavity as shown in Fig.~\ref{fig:cavity}(f). The isosurface reduces to a pair of points. The thickness of the cavity is $D=D_0=\lambda/2=\pi/n_0 k$. Also, assume $\theta_1=\theta_2$ and $\phi_1=\phi_2=\pi/2$. Then we have
\begin{equation}\label{eq:fundamental_interaction_rad}
\Gamma(\omega_0,\mathbf{R})
=
\frac{3}{2} n_0 \Gamma_0 \cos (k_z \Delta z),
\end{equation}
where $\Gamma_0 = d_1 d_2 \omega^3/ 3\pi \epsilon_0 c^3 $ is the spontaneous decay rate in vacuum.

%\begin{figure}
%\begin{center}
%\epsfig{figure=sfig_1.eps,width=\linewidth}
%\end{center}
%\caption{(a) Schematic of a Fabry$-$P\'{e}rot cavity with a medium sandwiched by two PEC plates.  }
%\label{fig:schematic_cavity}
%\end{figure}

\begin{figure*}
\begin{center}
\epsfig{figure=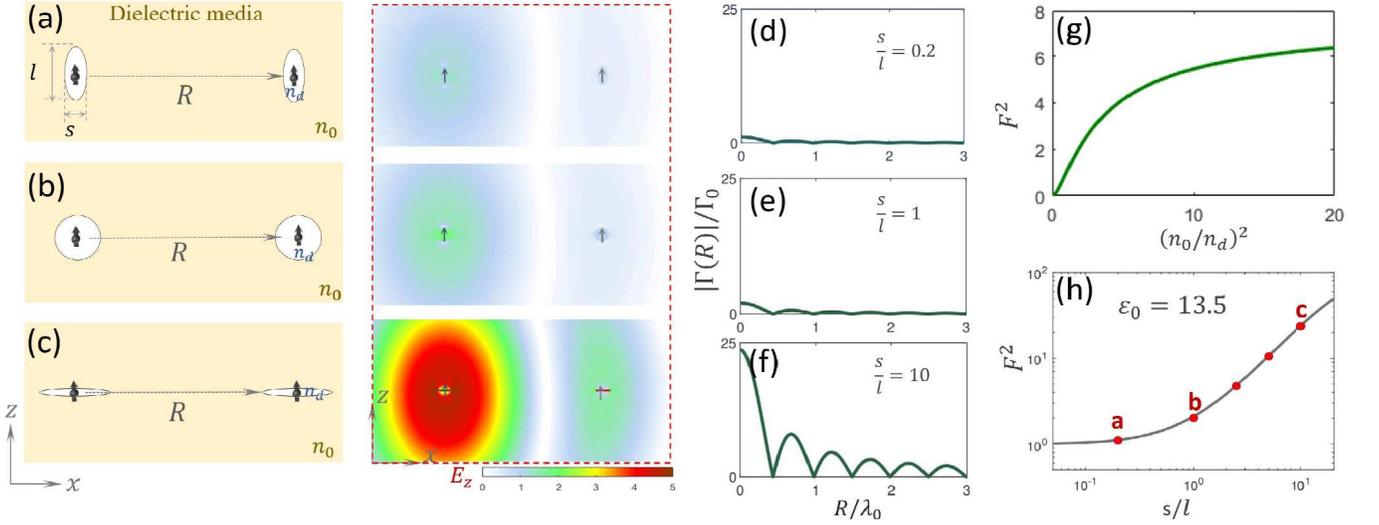,width=\linewidth}
\end{center}
\caption{(a-c) Left: Schematics of an isotropic dielectric media ($\epsilon_0=13.5$) with different geometrical air defects. Each defect is doped by a QT with a fixed dipolar orientation $[0, 0, 1]$.   The diameter ratios of the elliptical defects are $l/s=$ (a) $0.2$, (b) $1$, and (c) $10$, respectively. Right: Corresponding electric field $E_z(\mathbf{r})$ on $y=0$ plane. (d-f) The absolution value of radiative interaction as a function inter-QT distance $R$ for the cases in (a-c), respectively. (g) The square enhancement factor $F^2$ as a function of index ration $(n_0/n_d)^2$ . (h) The square enhancement factor $F^2$ as a function of geometrical ration $l/s$. The grey solid curve is from theoretical estimation and the red solid dots are obtained from full-wave simulations. }
\label{fig:enhancement}
\end{figure*}

{\bf The interaction via evanescent modes ---}
The interaction includes real and imaginary parts in Eq.~(\ref{eq:interaction_expression}). We already have the imaginary part in Eq.~(\ref{eq:interaction_expression}). In the opposite, the real part of the interaction is responsible for the evanescent modes. From Eq.~(\ref{eq:interaction_expression}), we have
\begin{equation}
\Delta(\omega_0)  =  \frac{1}{\pi}\int_0^\infty d\omega
\left(
\frac{\Gamma(\omega)}{\omega-\omega_0}
+ \frac{\Gamma^\ast(\omega)}{\omega+\omega_0}
 \right).
\end{equation}
Then, the interaction via evanescent modes and radiative modes satisfies the Kramers-Kronig relationship. The evanescent interaction scaling with the inter-QT distance is same as the radiative case, but their phases are opposite. For the fundamental mode of a perfect cavity, the interaction via evanescent modes is diverging while the radiative interaction is a constant as shown in Eq.~(\ref{eq:fundamental_interaction_rad}).

\section{Local field enhancement}\label{app:enhancement}

Our proposal to enhance interaction strength is based on a low-refractive-index defect, which only enhances the local field without any other negtive influence on the interaction range.

We consider an air defect with a radius $s$ along $x$ axis and two radiuses $l$ along $y,z$ axises in Fig. 4 (a) and its induced dipole moment along $x$ axis is given by
\begin{equation}
P =
\epsilon_0 n_0^2 n_d^2 \alpha E_{\mathrm{0},i}(\mathbf{r}) .
\end{equation}
Then, the polarizability is written as
\begin{equation}
\alpha = \frac{4}{3}\pi s l^2 \frac{n_\mathrm{d}^2 -n_0^2}{ n_0^2 +  r \left( n_\mathrm{d}^2 - n_0^2 \right)},
\end{equation}
where the geometrical factor is $r=(s l^2)/2 \int_0^\infty dq/(q+s^2)f(q) $ with $f(q)=\sqrt{(q+s^2 )(q+l^2 )^2}$.
Since we consider the size of the defect is much smaller than the wavelength  ($d_i\ll \lambda$), the electric field is smooth inside the defect. Thus, we have the equation as
\begin{equation}
\begin{split}
P
= &
\epsilon_0 n_d^2(n_d^2 - n_0^2) \iiint_{V_\mathrm{d}} E_{\mathrm{d},x}(\mathbf{r}) d^3\mathbf{r} \\
\approx &
\frac{4}{3}\pi sl^2 \epsilon_0 n_\mathrm{d}^2(n_\mathrm{d}^2 - n_0^2)  E_{\mathrm{d},x},
\end{split}
\end{equation}
where $V_\mathrm{d}$ is the volume of the ellipsoid defect and $E_{\mathrm{d},x}$ is $x$ component of electric field inside the defect.
Thus, the enhancement factor along $x$ axis is given by
\begin{equation}
F^2 =\frac{|E_{\mathrm{d},x}|^2}{|E_{0,x}|^2} =\Big[ \frac{n_0^2}{\left(1-r \right) n_0^2+r n_\mathrm{d}^2 }  \Big] ^2,
\end{equation}
where $n_\mathrm{d}$ and $n_0$ are the dielectric constants of defect and background, respectively. $r$ is the geometrical factor of the defect normal to $y$-$z$ plane. For a spherical geometry, i.e. $l=s$, the geometrical factor is $r=1/3$ and then the enhancement factor is $F^2=[3n_0^2/(2n_0^2+n_\mathrm{d}^2 )]^2=2.2$. If $s \ll l$, the enhancement factor will be greatly enhanced. For example, in Fig.~\ref{fig:enhancement} (c) $F=25$ with the diameter ratio $s=0.1 l$.

For a two-dimensional case, the ellipsoid is reduced to an elliptical defect. Then, the geometrical factor becomes $r=(ls)/2 \int_0^\infty dq/(q+s^2)f(q) $ with $f(q)=\sqrt{(q+l^2 )(q+s^2 )}$.

Then, two dipoles are placed inside the air defects with the dipole orientation parallel to the $z$ axis. The interaction then can be written as
\begin{equation}
\Gamma_\mathrm{en} (R)=F^2 \Gamma(R),
\end{equation}
where $\Gamma(R)$ is the radiative interaction without low-index defects. We show the radiative interaction as a function separation distance for different geometrical defects in Fig.~\ref{fig:enhancement} (b-f). Also, in Fig~\ref{fig:enhancement} (h) and (g), we plot the interaction strength as a function of dielectric ratio and geometrical ratio, respectively. As we can see, the enhancement of interaction strength does not depend on any other range-related parameters.

We also do full-wave simulation for the local field enhancement by solving the Maxwell's equations. As shown in Fig.~\ref{fig:enhancement}, two air defects are placed in a high-index material. In the simulation, we fix the background index as $n_0=3.67$ and the long radius of the ellipsoid defect is $l=0.025\lambda_0$. The short one $s$ is tuned as $0.125$, $0.025$, and $0.025\lambda_0$.  The left point dipole is set as the source and then we can get the electric field at the other dipole location. The electric field calculated from the full-wave simulation matches the theoretical estimations, as shown in Fig.~\ref{fig:enhancement} (h).

\section{Summary of interaction in various photonic environments}\label{app:photonics}

In this section, we will introduce the detailed information about the interaction range and interaction strength via propagating modes in Fig. 1 (c).

{\bf Uniform media and Index-near-zero metamaterials ---}
In isotropic uniform media, the dispersion relation is given by $\omega = (c/n)k_0 = c k$ with $n$ is the refractive index of the medium. The interaction via propagating modes is written as
\begin{equation}
\begin{split}
\Gamma(R,n) = \frac{d_1d_2k^3}{4\pi\epsilon_0 n^2}\Big[(\delta_{12} -  \hat{R}_1\hat{R}_2)  \frac{\sin{kR}}{kR}  \\
+ (\delta_{12} -  3\hat{R}_1\hat{R}_2)  \Big( \frac{\cos{kR}}{(kR)^2} -  \frac{\sin{kR}}{(kR)^3}  \Big)
\Big],
\end{split}
\end{equation}
where $\delta_{12}=\hat{\mathbf{d}}_1 \cdot \hat{\mathbf{d}}_2$ and $\hat{R}_{1,2}=\hat{\mathbf{d}}_{1,2}\cdot \hat{\mathbf{R}}$. The interaction strength is $\Gamma(R=0,n) = n\Gamma_0$ and the interaction range is $\ell = \ell_0/n$, where $\Gamma_0 = d_1 d_2 k_0^3/(3\pi \epsilon_0)$ is the spontaneous decay rate in vacuum and $\ell_0\approx 0.6\lambda_0 $ is the interaction range in vacuum. We plot the uniform medium with refractive index from $n=1.5$ to $3$ and the index-near-zero material~\cite{FA:2013,LR:2017} with effective refractive index from $n=0.01$ to $0.3$ in Fig. 1, respectively. The dipole orientations $\hat{\mathbf{d}}_{1,2}$ of quantum transitions are same and they are vertical to $\mathbf{R}$.

{\bf Bandedge of Photonic crystals ---}
The dispersion relation of the period dielectric photonic crystal is written as \cite{JW:1991}
\begin{equation}
\omega = \frac{c}{4na_n} \mathrm{arccos} \Big[ \frac{4n\cos(ka)+(n-1)^2}{(n+1)^2}  \Big]
\end{equation}
with the condition of $(n+1)a_{n}=a/2$, where $a$ is the lattice constant and $a_n$ is the radius of dielectric region in a unit cell. Then, we have the size of isosurface and the group velocity for a fix frequency from the dispersion relation. Then, we substitute these quantities into Eq. (9). The data in Fig. 1 (c) is plotted with the refractive index from $n=1.5$ to $3$ and the frequency near the band edge.

{\bf Hyperbolic meta-materials ---}
The dispersion relation of the hyperbolic material~\cite{CJ:2017} is $k_x^2/\epsilon_z+k_z^2/\epsilon_x=\omega^2/c^2$ with $\epsilon_x\epsilon_z<0$. The interaction between quantum transitions can be calculated from the dyadic Green's function. Because the interaction via propagating modes is diverging at $R=0$, the interaction strength is infinite and the interaction range is zero. We assume the interaction strength is $\Gamma_m=\Gamma(R=0.1\lambda)$ and the interaction range is $\ell = R(\Gamma=\Gamma_m/2)-0.1\lambda$. The data in Fig. 1 is based on two sets of parameters $\mathrm{Re}[\epsilon_x]=-2\mathrm{Re}[\epsilon_z]=2$ and $\mathrm[Re][\epsilon_x]=-\mathrm{Re}[\epsilon_z]=1$ with imaginary part of dielectric constant from $\mathrm{Im} [\epsilon_x]=\mathrm{Im} [\epsilon_z]=0.01$ to $0.2$.

{\bf Plasmonic plates ---}
The electric field of air-silver interface \cite{ZLL:2011} is $E_\mathrm{spp} \sim \eta R^{-1/2} e^{(ik_\mathrm{spp}-\alpha )R} E_0$, where $E_0$ is the magnitude of incident dipole radiating field, $\eta$ is the coupling efficiency between the quantum transition and the surface plasmon polariton (SPP) modes, $k_\mathrm{spp}=\mathrm{Re} \big[k_0\sqrt{\epsilon_0\epsilon_\mathrm{Ag}/(\epsilon_0\epsilon_\mathrm{Ag})}\big]$ is the SPP wave vector and $\alpha=\mathrm{Im} \big[k_0\sqrt{\epsilon_0\epsilon_\mathrm{Ag}/(\epsilon_0\epsilon_\mathrm{Ag})}\big]$.
$\epsilon_\mathrm{Ag}$ is the permittivity of silver. The interaction via propagating modes is given by
\begin{equation}
\Gamma(R)= \frac{E_\mathrm{spp}^2}{E_\mathrm{vacuum}^2}\Gamma_0 \approx \eta^2 \Gamma_0 e^{-2\alpha R} R \frac{\sin^2(k_\mathrm{spp} R) } {\sin^2 (k_0 R)}.
\end{equation}
In Fig. 1 (c), the parameters are $\lambda=400$, $450$, $500$, and $700[\mathrm{nm}]$; $\eta(0.45\eta)=0.44$, $0.29$, $0.22$, and $0.115$;  $k_\mathrm{spp} /k_0=1.131$, $1.076$, $1.053$, and $1.021$;  $\alpha=56.183$, $18.13$, $7.905$, and $1.4833 \ [\times10^{-5}\mathrm{nm}^{-1}]$, respectively.

{\bf Weyl photonic crystals ---}
In the Weyl photonic crystal, the interaction strength and the interaction range via propagating modes are from the data in Ref.~\cite{YZM:2019}. The dielectric constant of the double-gyroid structure is $\epsilon=13$.

\section{Numerical method of Green's function}\label{app:num}

The cavity can be considered as a multi-layer structure. Then, the interaction between two QTs in a cavity is given by
\begin{equation}
\begin{split}
\Sigma_{12}=  \frac{\omega^2}{\hbar \epsilon_0 c^2}
\mathbf{d}_2 ^{\dagger}\cdot \bar{\bar{\mathbf{G}}}(\mathbf{r}_2,\mathbf{r}_1;\omega) \cdot \mathbf{d}_1
\end{split}
\end{equation}
where $\bar{\bar{\mathbf{G}}}(\mathbf{r}_2,\mathbf{r}_1;\omega)$ is the dyadic Green's function. Its matrix form is written as
\begin{equation}\label{eq:gf_matrix}
\bar{\bar{\mathbf{G}}}  \left( \mathbf{r}_2,\mathbf{r}_1;\omega \right)  =
\begin{pmatrix}
G_{xx} & G_{xy} & G_{xz} \\
G_{yx} & G_{yy} & G_{yz} \\
G_{zx} & G_{zy} & G_{zz}
\end{pmatrix} ,
\end{equation}
where the element is the $j$-component electric field at $\mathbf{r}_2$ with a point source with dipole orientation $i$ at $\mathbf{r}_1$ and $i,j=x,y,z$.

{\bf Numerical method in multi-layered structures (Fabry$-$P\'{e}rot cavities) ---}
For a multilayered structure, the dyadic Green's function is given by~\cite{MM:1997}
\begin{equation}
G_{ij} = \frac{1}{2\pi} \int_0^\infty g_{ij}J_n(k_\parallel \parallel) k_\parallel d k_\parallel
\end{equation}
with $i,j=x,y,z$. $J_n(x)$ is the Bessel function of the first kind of order $n$.
\begin{equation}
\begin{split}
\bar{\bar{\mathbf{g}}}  \left( \mathbf{r}_2,\mathbf{r}_1;\omega \right)
= &
\begin{pmatrix}
g_{xx} & g_{xy} & g_{xz} \\
g_{yx} & g_{yy} & g_{yz} \\
g_{zx} & g_{zy} & g_{zz}
\end{pmatrix}
\\
 = &
- \hat{\mathbf{u}}^T \hat{\mathbf{u}} V_i^e
- \hat{\mathbf{v}}^T \hat{\mathbf{v}} V_i^h \\
   &
+ \hat{\mathbf{z}}^T \hat{\mathbf{u}} \frac{k_\parallel}{\omega\epsilon_0 \epsilon_{z_2}} I_i^e
+ \hat{\mathbf{u}}^T \hat{\mathbf{z}} \frac{k_\parallel}{\omega\epsilon_0 \epsilon_{z_1}} V_v^e \\
&
+\hat{\mathbf{z}}^T \hat{\mathbf{z}} \frac{1}{i\omega\epsilon_0 \epsilon_{z_1}} I_i^e
\left[
\frac{k_\parallel^2}{i\omega\epsilon_0\epsilon_{z_2}} I_v^e
-\delta \left(z_2-z_1\right)
\right],
\end{split}
\end{equation}
where
\begin{equation}
\begin{split}
\hat{\mathbf{u}}
&=  +\frac{k_x}{k_\parallel}\hat{\mathbf{x}}  +\frac{k_y}{k_\parallel}\hat{\mathbf{y}}, \\
\hat{\mathbf{v}}
&=  -\frac{k_x}{k_\parallel}\hat{\mathbf{x}}  +\frac{k_y}{k_\parallel}\hat{\mathbf{y}}, \\
k_\parallel
&=
|\mathbf{k}_\parallel|
= \sqrt{k^2-k_z^2}.
\end{split}
\end{equation}
The transmission-line Green's functions are given by
\begin{equation}
\begin{split}
V_i^p
&=  \frac{Z_n^p}{2}\left(
A_{n0}^p + \frac{ A_{n1}^p + A_{n2}^p +A_{n3}^p + A_{n4}^p }{D_n^p} \right), \\
V_v^p
&= \frac{1}{2}\left(
A_{n0}^p + \frac{ A_{n1}^p - A_{n2}^p +A_{n3}^p - A_{n4}^p }{D_n^p}  \right), \\
I_i^p
&= \frac{1}{2}\left(
A_{n0}^p + \frac{ -A_{n1}^p + A_{n2}^p +A_{n3}^p - A_{n4}^p }{D_n^p} \right), \\
I_v^p
&= \frac{1}{2 Z_n^p}\left(
A_{n0}^p + \frac{ -A_{n1}^p - A_{n2}^p +A_{n3}^p +A_{n4}^p }{D_n^p} \right),
\end{split}
\end{equation}
where
\begin{equation}
\begin{split}
Z_n^e
&= \frac{k_{zn}}{\omega\epsilon_n}, \ \
Z_n^e = \frac{\omega \mu_n}{k_{zn}},   \\
D_n^p &= 1-\Gamma_{n,n+1}^p  \Gamma_{n,n-1}^p ,\\
A_{ns}^p
&= a_{ns}^p e^{-ik_{zn} \gamma_s}
\end{split}
\end{equation}
with $s=0,1,2,3,4$.
The coefficients are
\begin{equation}
\begin{split}
a_0^p &= 1, \\
a_1^p &= \Gamma_{n,n-1}^p, \\
 a_2^p &= \Gamma_{n,n+1}^p, \\
a_3^p &= a_4^p =a_1^p a_2^p,
\end{split}
\begin{split}
\ \ \ \ \ \  \gamma_0 & = |z_2 -z_1|,  \\
\ \ \ \ \ \  \gamma_1 & = 2z_n-\left(z_2+z_1\right),  \\
\ \ \ \ \ \  \gamma_2 & = \left( z_2 + z_1 \right) - 2 z_{n-1},  \\
\ \ \ \ \ \ \gamma_3 & = 2 t_n -\gamma_0,  \\
\ \ \ \ \ \ \gamma_4 & = 2 t_n  + \gamma_0.
\end{split}
\end{equation}
The generalized reflection coefficients are given by
\begin{equation}
\begin{split}
\Gamma_{n,n+1}^p &= \frac{R_{n,n+1}^p + \Gamma_{n+1,n+2}^p e^{-2i k_{zn+1}t_{n+1}}   }
{1+ R_{n,n+1}^p  \Gamma_{n+1,n+2}^p e^{-2i k_{zn+1}t_{n+1}}  } , \\
\Gamma_{n,n-1}^p &= \frac{R_{n,n-1}^p + \Gamma_{n-1,n-2}^p e^{-2i k_{zn-1}t_{n-1}}   }
{1+ R_{n,n-1}^p  \Gamma_{n-1,n-2}^p e^{-2i k_{zn-1}t_{n-1}}  } .
\end{split}
\end{equation}
The Fresnel reflection coefficients are given by
\begin{equation}
\begin{split}
R_{n,n+1}^p &= \frac{R_{n,n+1}^p + \Gamma_{n+1,n+2}^p e^{-2i k_{zn+1}t_{n+1}}   }
{1+ R_{n,n+1}^p  \Gamma_{n+1,n+2}^p e^{-2i k_{zn+1}t_{n+1}}  } , \\
R_{n,n-1}^p &= \frac{R_{n,n-1}^p + \Gamma_{n-1,n-2}^p e^{-2i k_{zn-1}t_{n-1}}   }
{1+ R_{n,n-1}^p  \Gamma_{n-1,n-2}^p e^{-2i k_{zn-1}t_{n-1}}  } .
\end{split}
\end{equation}

{\bf Full-wave simulation for dyadic Green's function ---}
For complex structures, the full-wave simulation is needed to calculate the dyadic Green's function. The matrix element of the dyadic Green's function is defined by
\begin{equation}
G_{ij}(\mathbf{r}_2,\mathbf{r}_1;\omega) = \frac{E_{ i}(\mathbf{r}_2 ,\omega) }{d_{1,j} (\mathbf{r}_1,\omega) },
\end{equation}
where $d_{1,j} (\mathbf{r}_1,\omega) $ is the source $j$-polarized dipole moment and $E_i$ is the $i$ component of electric field at $\mathbf{r}_2 $ due to the dipole radiation.

{\bf Interaction in Dirac photonic slabs ---}
We use a triangular lattice with two basis $\mathbf{a}_1 = [-1/2, \sqrt{3}/2]a$ and $\mathbf{a}_2 = [1/2,\sqrt{3}/2]a$. The dielectric constant of the slab is $8$ and hole radius is $0.45a$, where $a$ is the lattice constant.
Numerically, we use the MPB software package~\cite{johnson2001block} to calculate eigen modes of the Dirac photonic slab in Fig. 2. We set the resolution in the unit cell as $32 \times 32$. Then, the frequency of the Dirac points is $\omega_\mathrm{D}  \approx  0.862 [2\pi c/a]$, which falls between $4$th and $5$th bands.
Then we calculate the radiative interaction via propagating modes by the definition of Eq.~(\ref{eq:interaction_expression}).

\section{Non-Markovian dynamics between two quantum transitions}\label{app:dynamics}

\begin{figure}
\begin{center}
\epsfig{figure=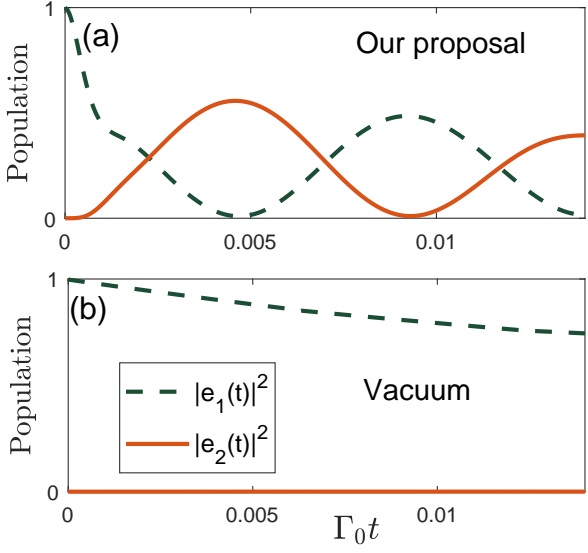,width=\linewidth}
\end{center}
\caption{Non-Markovian dynamics of the quantum transtions $1$ and $2$ in (a) our proposed structure with cavity and defect and (b) vacuum. The inter-distance $R$ is $0.5\lambda_0/n_0$ and the reflectance of the cavity mirror is $0.99996$. The enhancement factor is fixed as $F^2=80$ and the Bohr frequency is $\omega_B=10\omega_0$, the spontaneous decay rate at resonant frequency is $\Gamma_0 = \Gamma(\omega_0,R=0)=\omega_0/20000$.   }
\label{fig:dynamics}
\end{figure}

In this section, we introduce the dynamics between two quantum transitions in the non-Markovian limit by solving the time-dependent Schrodinger equation
\begin{equation}
i\frac{\partial}{\partial t} |\Psi(t)\rangle =  \mathcal{H} |\Psi(t)\rangle  ,
\end{equation}
where the Hamiltonian $\mathcal{H}$ has been given by Eq.~(\ref{eq:hamiltonian_tot}) and the wavefunction is
\begin{equation}
|\Psi(t) \rangle=
 \Big[
 e_1(t)\mathcal{S}_1^\dag + e_2(t)\mathcal{S}_2^\dag  +   \sum_\alpha c_\alpha(t) \mathcal{A}_\alpha^\dag
  \Big]
  |g_1,g_2;0\rangle.
\end{equation}
The propability amplitudes of  quantum transitions $1$ and $2$ are given by
\begin{equation}
\begin{split}
 \frac{d}{dt} e_1(t) = - i\omega_0 e_1(t) - \sum_{j=1,2} \int_0^\infty G_{1j}(t-\tau) e_j(\tau) d\tau, \\
 \frac{d}{dt} e_2(t) = - i\omega_0 e_2(t) - \sum_{j=1,2} \int_0^\infty G_{2j}(t-\tau) e_j(\tau) d\tau,
\end{split}
\end{equation}
where the memory kernel is written as
\begin{equation}
G_{jm}(t-\tau) = \int_0^\infty  \rho(\omega)\Gamma_{jm}(\omega) e^{-i\omega(t-\tau)} d\omega.
\end{equation}
Here, $\rho(\omega) =   1/[1+(\omega/\omega_B)^2]^4$ is the cutoff function due to the limit of Bohr frequency~\cite{KK:2000} $\omega_B$, and $\Gamma_{jm}(\omega)=\Gamma(\mathbf{r}_m-\mathbf{r}_2,\omega)$ with $j,m=1,2$. We consider a pair of quantum transitions with a separation distance of $\lambda_0/2n_0$. Fig.~\ref{fig:dynamics} shows the populations of quantum transitions $1$ and $2$ as a function of time for our proposed structure of a cavity doped with low-index defects, cavity, and vacuum, respectively. For the cavity with field-enhanced defect, the strong-and-long interaction results in fast oscillation between two quantum transitions.

%\bibliographystyle{apsrev4-2}
%\bibliography{longstrong_ref}

%

\end{document}